\documentclass[11pt,a4paper]{article} 
\usepackage{epsfig}
\usepackage{psfig}
\usepackage{cite}
\def\3{\ss}

\begin{document}

\title {\begin{flushright}{\large DESY--99--057}  \end{flushright}
\vspace*{2cm}
\bf\LARGE Measurement of Dijet Photoproduction at High Transverse Energies at HERA }
\author{ZEUS Collaboration }
\date{}
\maketitle
\begin{abstract}
\noindent 
The cross section for dijet photoproduction at high transverse 
energies is presented as a function of the transverse 
energies and the pseudorapidities of the jets.
The measurement is performed using a sample of 
$ep$-interactions corresponding to an integrated luminosity of 
6.3 pb$^{-1}$, recorded by the ZEUS detector. 
Jets are defined by applying a $k_T$-clustering  
algorithm to the hadrons observed in the final state. 
The measured cross sections are compared to next-to-leading order QCD 
calculations. In a kinematic regime where theoretical uncertainties are 
expected to be small, the measured cross sections are higher than 
these calculations. 
\end{abstract}

\pagestyle{plain}
\thispagestyle{empty}
\clearpage
\pagenumbering{Roman}

\begin{center}                                                                                     
{                      \Large  The ZEUS Collaboration              }                               
\end{center}                                                                                       
  J.~Breitweg,                                                                                     
  S.~Chekanov,                                                                                     
  M.~Derrick,                                                                                      
  D.~Krakauer,                                                                                     
  S.~Magill,                                                                                       
  B.~Musgrave,                                                                                     
  A.~Pellegrino,                                                                                   
  J.~Repond,                                                                                       
  R.~Stanek,                                                                                       
  R.~Yoshida\\                                                                                     
 {\it Argonne National Laboratory, Argonne, IL, USA}~$^{p}$
% ...
\newpage                                                                                           
$^{\    1}$ now visiting scientist at DESY \\                                                      
% ...
$^{  36}$ partially supported by the Foundation for German-Russian Collaboration                   
DFG-RFBR \\ \hspace*{3.5mm} (grant no. 436 RUS 113/248/3 and no. 436 RUS 113/248/2)\\              
% ...
                                                           %
                                                           %
\newpage   
                                                           %
                                                           %
\begin{tabular}[h]{rp{14cm}}                                                                       
$^{a}$ &  supported by the Natural Sciences and Engineering Research                               
          Council of Canada (NSERC)  \\                                                            
$^{b}$ &  supported by the FCAR of Qu\'ebec, Canada  \\                                            
$^{c}$ &  supported by the German Federal Ministry for Education and                               
          Science, Research and Technology (BMBF), under contract                                  
          numbers 057BN19P, 057FR19P, 057HH19P, 057HH29P, 057SI75I \\                              
$^{d}$ &  supported by the MINERVA Gesellschaft f\"ur Forschung GmbH, the                          
German Israeli Foundation, and by the Israel Ministry of Science \\                                
$^{e}$ &  supported by the German-Israeli Foundation, the Israel Science                           
          Foundation, the U.S.-Israel Binational Science Foundation, and by                        
          the Israel Ministry of Science \\                                                        
$^{f}$ &  supported by the Italian National Institute for Nuclear Physics                          
          (INFN) \\                                                                                
$^{g}$ &  supported by the Japanese Ministry of Education, Science and                             
          Culture (the Monbusho) and its grants for Scientific Research \\                         
$^{h}$ &  supported by the Korean Ministry of Education and Korea Science                          
          and Engineering Foundation  \\                                                           
$^{i}$ &  supported by the Netherlands Foundation for Research on                                  
          Matter (FOM) \\                                                                          
$^{j}$ &  supported by the Polish State Committee for Scientific Research,                         
          grant No. 115/E-343/SPUB/P03/154/98, 2P03B03216, 2P03B04616,                             
          2P03B10412, 2P03B05315, and by the German Federal Ministry of                            
          Education and Science, Research and Technology (BMBF) \\                                 
$^{k}$ &  supported by the Polish State Committee for Scientific                                   
          Research (grant No. 2P03B08614 and 2P03B06116) \\                                        
$^{l}$ &  partially supported by the German Federal Ministry for                                   
          Education and Science, Research and Technology (BMBF)  \\                                
$^{m}$ &  supported by the Fund for Fundamental Research of Russian Ministry                       
          for Science and Edu\-cation and by the German Federal Ministry for                       
          Education and Science, Research and Technology (BMBF) \\                                 
$^{n}$ &  supported by the Spanish Ministry of Education                                           
          and Science through funds provided by CICYT \\                                           
$^{o}$ &  supported by the Particle Physics and                                                    
          Astronomy Research Council \\                                                            
$^{p}$ &  supported by the US Department of Energy \\                                              
$^{q}$ &  supported by the US National Science Foundation \\                                       
\end{tabular}                                                                                      
                                                           %
                                                           %
\newpage
\setcounter{page}{1}
\pagenumbering{arabic}

\section{Introduction} \label{intro}

In photoproduction at HERA a quasi real photon, emitted from 
the incoming positron, collides with the incoming proton. 
In leading order quantum chromodynamics (QCD), two 
processes contribute to the photoproduction of jets: 
the direct process, in which the photon couples directly 
to a parton in the proton, and the resolved process, in which the 
photon acts as a source of partons, one of which scatters from 
a parton in the proton. Beyond the leading order in QCD,  
direct and resolved processes are not distinctly separable.  

The cross section for jet photoproduction is sensitive to the 
partonic structures of both the proton and the photon.
In the kinematic regime of the measurement 
presented in this paper, the fractional momentum $x$ at which partons 
inside the proton are probed
lies predominantly in the region 
between $10^{-2}$ and $10^{-1}$. 
At these $x$ values the parton densities in 
the proton are strongly constrained by measurements of the structure 
function $F_2^p$ in deep inelastic lepton-proton scattering~\cite{f2_proton}.
The fractional momentum $x_\gamma$ at which partons 
in the photon are probed lies between 0.1 and 1. 
For $x_\gamma$ values above $0.5$ the quark densities in the photon are 
not strongly constrained by $F_2^\gamma$ data obtained from 
$\gamma\gamma^*$ scattering at $e^+e^-$ experiments~\cite{f2_photon}.

The investigation presented in this paper aims to  
constrain more tightly the parton densities in the photon at high $x_\gamma$, 
where the contribution from quarks dominates,  
by exploiting their influence on the dijet photoproduction cross 
section. For this purpose the dijet cross section is measured in 
a kinematic regime where next-to-leading order (NLO) QCD calculations 
are expected to describe the data. 
It should be noted here that jet measurements at the 
Tevatron~\cite{tevatron}, although generally in good agreement with 
NLO-QCD, show discrepancies in the comparison of the 630~GeV and 
1800~GeV data sets. These may be connected to non-perturbative effects, 
such as a possible underlying event~\cite{tev_nlocomp}. A number of 
these effects, which may also be of relevance to the present study, 
have been investigated in this paper.

This paper builds on  
the improved understanding of jet photoproduction and of comparisons to 
NLO-QCD calculations gained in previous 
analyses~\cite{zeus_incl1}~-~\cite{zeus_multi} and on
a significant theoretical effort in the recent 
past~\cite{kks}~-~\cite{nloho2}. 
Events with two or more high-transverse-energy 
jets are used, one of which is required to have transverse energy greater 
than 14 GeV and the second one greater than 11 GeV. A previous jet 
photoproduction 
analysis~\cite{zeus_dij2} has shown that for jets with transverse energy greater than $11$~GeV, the dijet cross section agrees with NLO-QCD predictions, within the experimental uncertainties of that analysis.

\section{Experimental setup} \label{experiment}
The data used in this paper were collected in 1995 with the ZEUS
detector at HERA, colliding positrons at an energy of $E_e=27.5$~GeV 
with protons at an energy of $E_p=820$~GeV, yielding a total CM energy of
 $\sqrt{s}=\sqrt{4E_e E_p}\approx300$~GeV. The data sample corresponds
 an integrated luminosity of $6.3$~pb$^{-1}$.

The ZEUS detector is described in detail elsewhere~\cite{zeus_status}.
The main components used in this analysis are the uranium-scintillator 
calorimeter (CAL) and the central tracking detector (CTD). 
The CAL~\cite{zeus_status,cal} covers 99.9\% of the total solid angle 
and is subdivided into forward, barrel and rear parts, covering 
the pseudorapidity regions $4.3\ge\eta>1.1$, $1.1\ge\eta>-0.75$ 
and $-0.75\ge\eta>-3.8$, respectively\footnote{The ZEUS coordinate 
system is defined as right-handed with the $Z$ axis pointing in the proton 
beam direction, hereafter referred to as forward, and the $X$ axis horizontal, 
pointing towards the centre of HERA. The pseudorapidity is defined as 
$\eta=-\ln(\tan\frac{\theta}{2})$, where the polar angle $\theta$ is taken 
with respect to the proton beam direction.}.  
Test beam measurements yield energy 
resolutions of $\sigma(E)/E=18\%/\sqrt{E({\rm GeV})}$ for electrons and $\sigma(E)/E=35\%/\sqrt{E({\rm GeV})}$ for 
hadrons~\cite{caltest}.
The CTD~\cite{ctd} is a cylindrical drift 
chamber, situated in a 1.43 T solenoidal magnetic field, covering the polar angular region $15^\circ  < \theta < 164^\circ $.
The transverse momentum resolution for full-length tracks can be 
parametrised as
$\sigma(p_T)/p_T=0.0058p_T \oplus 0.0065 \oplus 0.0014/p_T$, with $p_T$ in GeV.
The luminosity collected by ZEUS is measured from the rate of the Bremsstrahlung process $e^+p\rightarrow e^+p\gamma$. A three-level trigger system is used to select events online~\cite{zeus_dij2,zeus_status}.

\section{Definition of the cross section}\label{definition}
The relevant variables for the dijet cross section measurement presented in this paper are the following:

\begin{itemize}
\item the transverse energy, $E_T^{jet}$, the azimuthal angle, $\phi^{jet}$, and the pseudorapidity, $\eta^{jet}$,
of the jets;
\item the scaled energy transfer from the positron to the proton in the proton's rest frame, defined as: 
\begin{equation}y= \frac{q\cdot p}{k\cdot p}\,,\,\,0<y<1\,,\end{equation}
where $q$, $k$ and $p$ are the four-momenta of the exchanged photon, the 
incoming positron and the incoming proton, respectively. 
Neglecting mass terms, $y$ is related 
to the centre-of-mass energy in the photon-proton system, 
$W_{\gamma p}=\sqrt{ys}$. In the photoproduction regime, 
where the exchanged photon is almost real, $y$ is equivalent to the 
fractional energy of the incoming positron carried by the photon; 
\item the fractional longitudinal momentum of the photon 
participating in the production of the two highest-transverse-energy 
jets, defined as~\cite{zeus_dij1}:
\begin{equation}x_\gamma^{obs}=\frac{E_{T\,1}^{jet}e^{-\eta_1^{jet}}+E_{T\,2}^{jet}e^{-\eta_2^{jet}}}{2yE_e}\,,\label{form:xg}\end{equation}
where $E_{T\,1,2}^{jet}$ and $\eta_{1,2}^{jet}$ are the transverse energies and the pseudorapidities of the two highest-transverse-energy jets;
\item the virtuality of the exchanged photon:
\begin{equation}Q^2=-q^2\,.\end{equation} 
\end{itemize}

The cross section presented in this paper is compared to NLO-QCD
predictions. 
It is restricted to a specific set of conditions, 
to minimise theoretical uncertainties.
\begin{itemize}
\item An asymmetric cut is applied on the transverse energy of the two 
  highest-transverse-energy jets. The application of a symmetric cut poses 
a stability problem for some of the available NLO-QCD 
calculations~\cite{symmcut1,symmcut2}.
\item Symmetrisation of the cross section with respect to the pseudorapidity 
of the two highest-transverse-energy jets has been claimed to remove infrared 
instabilities in the NLO-QCD calculations~\cite{ggk}. This entails analysing 
each event twice, as explained below.
\item Jets are defined using the longitudinally invariant $k_T$-clustering  
algorithm~\cite{kt1} in the inclusive mode~\cite{kt2}, where the 
parameter $R$ is chosen equal to 1. This 
algorithm provides a jet reconstruction that is suitable for 
comparisons between data and theory~\cite{jetalgorithms}.
\end{itemize}

The dijet photoproduction cross section presented in this paper 
refers to events in which at least two jets, as defined by the 
$k_T$-clustering  algorithm, are found in the 
hadronic final state. These jets are required to have pseudorapidities 
between $-1$ and $2$, transverse energy of the highest-transverse-energy jet, 
$E_{T\,leading}^{jet}$, greater than 14 GeV and the transverse energy of the 
second-highest-transverse-energy jet, $E_{T\,second}^{jet}$, greater 
than 11 GeV. The cross section is given in the kinematic region 
defined by: $Q^2<1$~GeV$^2$ and $0.20<y<0.85$.   

This cross section is measured as a function of three variables: 
$E_{T\,leading}^{jet}$, $\eta^{jet}_1$ and $\eta^{jet}_2$. 
The cross section is symmetrised with respect to the pseudorapidities 
of the two jets. Every event contributes twice to the 
cross section, once with $\eta^{jet}_1=\eta^{jet}_{leading}$ and 
$\eta^{jet}_2=\eta^{jet}_{second}$ and a second time with 
$\eta^{jet}_1=\eta^{jet}_{second}$ and $\eta^{jet}_2=\eta^{jet}_{leading}$.

The cross section is determined for the full range of 
$x_\gamma^{obs}$ values and for 
a direct-photoproduction-enriched region with $x_\gamma^{obs}>0.75$.
The cross section as a function of the pseudorapidity of the jets is also 
measured in a narrower band of $y$ values between 0.50 and 0.85, where 
the sensitivity to the photon structure is expected to be higher, as will be explained in section \ref{highy}.

\section{Comparisons to NLO-QCD}\label{nloqcd}

The measured cross sections are compared to NLO-QCD calculations by 
four different groups: P. Aurenche et al.~\cite{nloabfg}, S. Frixione et al.~\cite{symmcut2,nlofr}, B. Harris et al.~\cite{nloho} and M. Klasen et al.~\cite{nlokk}.
These calculations differ in the handling of 
divergences~\cite{phasespaceslicing,subtr}.

All calculations use the CTEQ4M~\cite{cteq4m} parameterisation of the
parton densities in the proton. The value of $\Lambda_{QCD}$ is 
chosen to match that of this set of parton distribution functions.
For the parton densities in the photon 
three parameterisations are used, GRV-HO~\cite{grv1,grv2}, 
GS96-HO~\cite{gs96} and AFG-HO~\cite{afg}.

In all calculations the renormalisation and 
factorisation scales are chosen equal to the transverse energy of the 
highest-transverse-energy jet.
The variation in the NLO-QCD calculations of the presented cross 
section has been found to be less than $15\%$, when the scales are varied 
between half and twice this value.

The NLO-QCD calculations do not include fragmentation.
Jets are defined on the basis of the outgoing partons.  
While the momenta of jets at high transverse energies are expected to 
correspond closely to the 
momenta of the partons produced in the hard subprocess, the measured 
jet cross sections are affected at some level by the fragmentation.
In a study using the HERWIG 5.9 and the PYTHIA 5.7 Monte 
Carlo photoproduction models, the dijet cross section 
for jets of hadrons was compared to that for partons produced in the 
two-to-two hard subprocess and in the parton showers which were grouped 
into ``parton jets'' using the $k_T$-clustering algorithm.

In HERWIG the change in the cross section due to the fragmentation 
was found to be less than $10\%$ in most of the present kinematic region. 
However for events in which one jet has $\eta^{jet}<0$ the cross 
section is reduced by more than $10\%$ due to fragmentation and when 
both jets have $\eta^{jet}<0$ the cross section is reduced by $\sim 40\%$.
In PYTHIA the reduction of the cross section due to fragmentation is 
much smaller, but shows the same trend.
In a related study, 
presented in reference~\cite{nloho2}, HERWIG 5.9 was used to compare 
the cross section for jets of hadrons 
to that for jets of partons, produced in the two-to-two hard subprocess.
The relative 
difference between these cross sections was found to be less than 20\%, 
except again for events with backward jets where the reduction of the 
cross section due to fragmentation exceeds $20\%$ and is again largest 
($\sim 50\%$) when both jets have $\eta^{jet}<0$. 

Since the effect of fragmentation on the cross section depends on the 
Monte Carlo model, no attempt was made to correct the data for these effects.
Instead, the effect of fragmentation is considered as a 
theoretical uncertainty.

\section{Energy corrections}\label{kinvar}
Kinematic variables are reconstructed using a combination of track 
and calorimeter information that optimises the resolution 
of reconstructed kinematic variables~\cite{briskin}.
The selected calorimeter clusters and tracks are referred to as Energy Flow 
Objects (EFOs).  

The use of track information reduces the sensitivity to energy losses in 
inactive material in front of the CAL. However, the energy of particles 
for which no track information is available (e.g. because the energy is  
deposited by a neutral particle), must be measured using CAL information. 
These energies have to be corrected for the energy losses 
in the inactive material.
The conservation of energy and momentum in 
neutral-current deep inelastic scattering events is exploited 
to determine the required energy corrections~\cite{thesis} by 
balancing the scattered positron with the hadronic 
final state.
This is done for data and Monte Carlo event samples independently.
The EFOs thus corrected are used both for the reconstruction of jets 
and to determine kinematic variables. 
Comparisons between data and Monte Carlo of kinematic variables, 
reconstructed using corrected EFOs, lead to the assignment of a 3\% 
correlated systematic uncertainty and a 2\% uncorrelated systematic 
uncertainty in the transverse jet energies and in the hadronic 
variables~\cite{thesis}.

\section{Event selection} \label{evsel}

After applying the energy corrections described in section \ref{kinvar}, 
dijet events are selected from those events triggered by the dijet
trigger~\cite{zeus_dij2} using the following procedures and cuts 
designed to remove sources of background:

\begin{itemize}
\item The $k_T$-clustering algorithm, in the inclusive mode with $R=1$, is 
applied to the corrected EFOs. 
Events are selected in which at least two jets are found with: 
$-1<\eta^{jet}<2$, $E_{T\,leading}^{jet}>14$~GeV and 
$E_{T\,second}^{jet}>11$~GeV.
\item To remove background due to proton beam-gas interactions and cosmic 
showers, a cut is made on the longitudinal position of the reconstructed 
interaction vertex
\begin{equation}-40\,{\rm cm}\,<Z_{vertex}<40\,{\rm cm}\,.\end{equation}
\item To remove background due to charged-current deep inelastic scattering events, 
a cut is made on the relative missing transverse momentum: 
\begin{equation}\frac{P_T}{\sqrt{E_T}}<1.5\,\sqrt{{\rm GeV}},\end{equation}
where $P_T$ and $E_T$ are the transverse momentum and the transverse energy of the event, calculated on the basis of corrected EFOs.
\item The rejection of neutral-current deep inelastic scattering (NC-DIS) events is based on the variable $y$.
If a scattered positron candidate 
with energy greater than $5$~GeV is found in the calorimeter, $y$ can be 
calculated from the energy $E^\prime_e$ and the polar angle 
$\theta_e^\prime$ of this positron candidate using the formula: 
$y_{elec}=1-\frac{E^\prime_e}{2E_e}\left(1-\cos{\theta_e^\prime}\right)$. 
These events are rejected when: 
\begin{equation}y_{elec}<0.7\,.\end{equation}
The variable $y$ can also be reconstructed from the observed 
hadronic final state using the Jacquet-Blondel formula~\cite{jb}: 
$y_{JB}=\sum (E-p_z)/2E_e$, where the sum runs over all corrected EFOs. 
For all events it is required that 
\begin{equation}0.20<y_{JB}<0.85\,.\end{equation} 
This cut removes unidentified NC-DIS events, for which $y_{JB}$ peaks at 1, 
and proton beam-gas interactions, which mostly have low 
$y_{JB}$ values.
The cuts on $y_{elec}$ and $y_{JB}$ effectively restrict the range of 
the virtuality of the exchanged photon to $Q^2<1$~GeV$^2$, with a 
median of about $10^{-3}$~GeV$^2$.
\end{itemize}

After the application of the described selection criteria, a sample of 8690
events remain. The contamination of this sample due to background processes 
was found to be negligible.

\section{Event characteristics} \label{mi}

Photoproduction events, generated using Monte Carlo programs, are 
used for the determination of acceptance 
and migration corrections and for the study of systematic uncertainties.
These events are passed through a 
full simulation of the ZEUS detector and undergo the same 
energy-correction procedure as the data.

Two leading-order Monte Carlo programs were used to generate dijet 
photoproduction events, HERWIG 5.9~\cite{herwig1,herwig2} 
and PYTHIA 5.7~\cite{pythia1,pythia2}. Both models use leading-order 
matrix elements, but they differ in the treatment of 
parton showers, hadronisation and the virtuality spectrum of the 
exchanged photon.  
No additional process that would produce soft or hard underlying 
events is included in the simulations. 

Direct and resolved event samples are generated separately. 
The parton density functions used to generate 
both Monte Carlo samples are CTEQ3-LO~\cite{cteq3_lo} for the 
proton and GRV-LO~\cite{grv1,grv2} for the photon.

As the Monte Carlo models do not include higher order matrix elements, 
they are not expected to describe the absolute normalisation of the 
cross section. 
To obtain the best agreement between data and Monte Carlo, the 
normalisation of the direct and resolved contributions is determined 
from a fit to the measured $x_\gamma^{obs}$ spectrum. As a result 
the direct contribution of the HERWIG Monte Carlo is scaled by 
a factor 1.83 and the resolved contribution by a factor 1.72. For 
PYTHIA the direct contribution is scaled by 1.28 and the resolved 
contribution by 1.27. When these factors are applied, both Monte Carlo 
models are found to give a reasonable description of various distributions, 
such as the $\eta^{jet}$ and $E_T^{jet}$ spectra.  
 
\begin{figure}
\begin{center}
\epsfxsize = 5.in
\epsfysize = 5.in
\epsffile{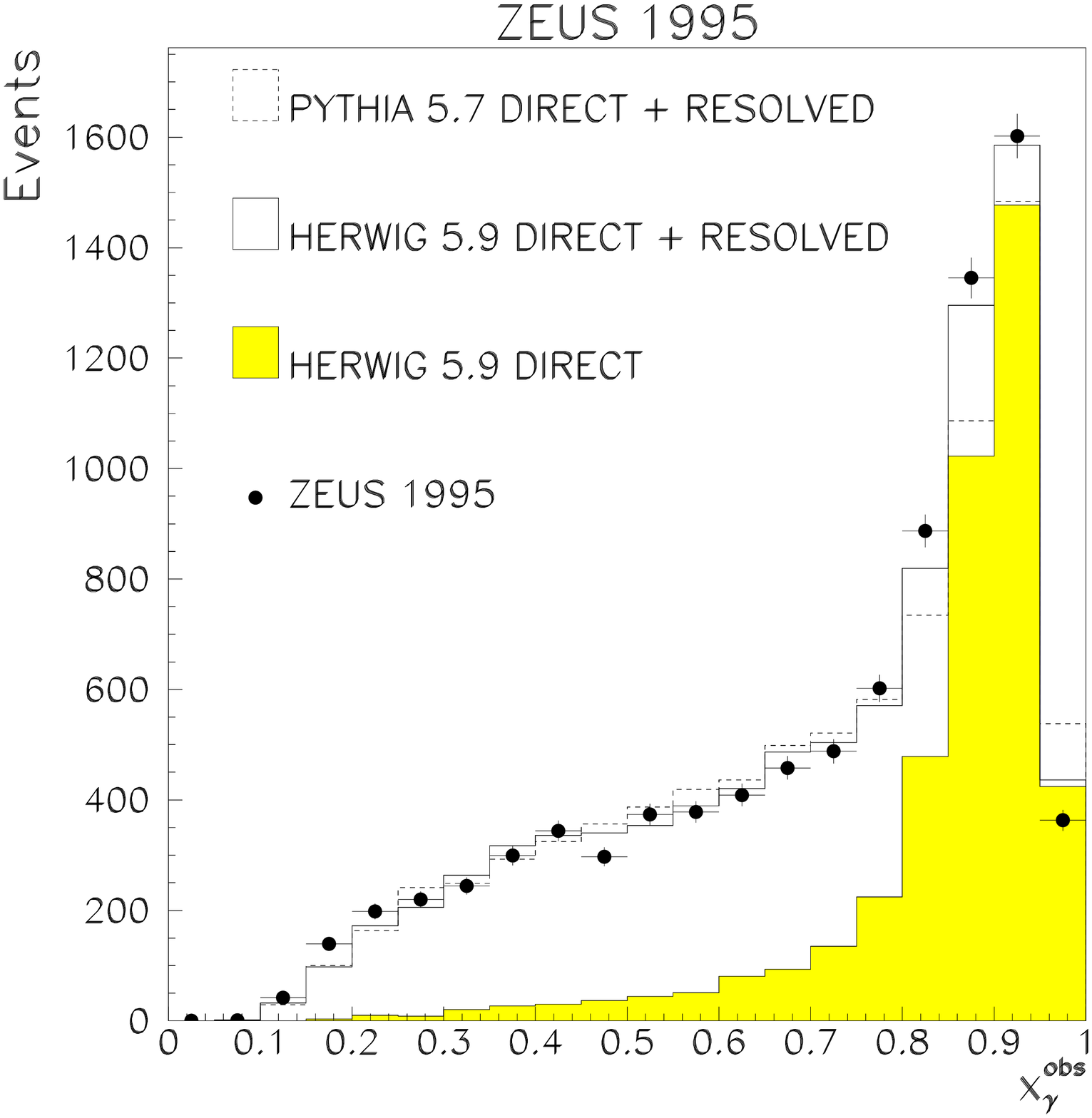}
\end{center}
\caption{\footnotesize The $x_\gamma^{obs}$ spectrum of the selected dijet sample, compared to the HERWIG 5.9 and the PYTHIA 5.7 Monte Carlo predictions, which have been weighted as described in section \ref{mi}. The direct component from the HERWIG Monte Carlo is shown separately as the shaded histogram. Only statistical uncertainties are plotted\label{fig:xg}}
\end{figure}

The $x_\gamma^{obs}$ spectrum for the selected 
sample of the 8690 dijet events is shown in Fig. \ref{fig:xg}, 
where $x_\gamma^{obs}$ is determined on the 
basis of corrected EFOs. The data show a clear peak near 
$x_\gamma^{obs}\sim 1$, attributed, at leading order to 
a predominance of ``direct'' events, and a tail 
towards low $x_\gamma^{obs}$ values, attributed to  
``resolved'' events. 
The data are compared to the HERWIG and PYTHIA Monte Carlo predictions,  
including the normalisation 
factors given above. The direct component of the HERWIG Monte Carlo 
is shown separately. The direct photoproduction events 
peak at high $x_\gamma^{obs}$ values. Therefore, selecting events with 
$x_\gamma^{obs}>0.75$ yields a sample strongly 
enriched with direct photoproduction events.
After application of the normalisation factors described above, the 
Monte Carlo predictions are in good agreement with the data. The shape of the 
peak at high $x_\gamma^{obs}$ is best described by the HERWIG Monte Carlo. 
Given the agreement in this distribution and in distributions like the 
$\eta^{jet}$ and $E_T^{jet}$ spectra (not shown here), 
the HERWIG Monte Carlo sample is used to 
determine acceptance and migration corrections and to study 
systematic uncertainties.

\begin{figure}
\begin{center}
\epsfxsize = 5.4in
\epsfysize = 5.5in
\epsffile{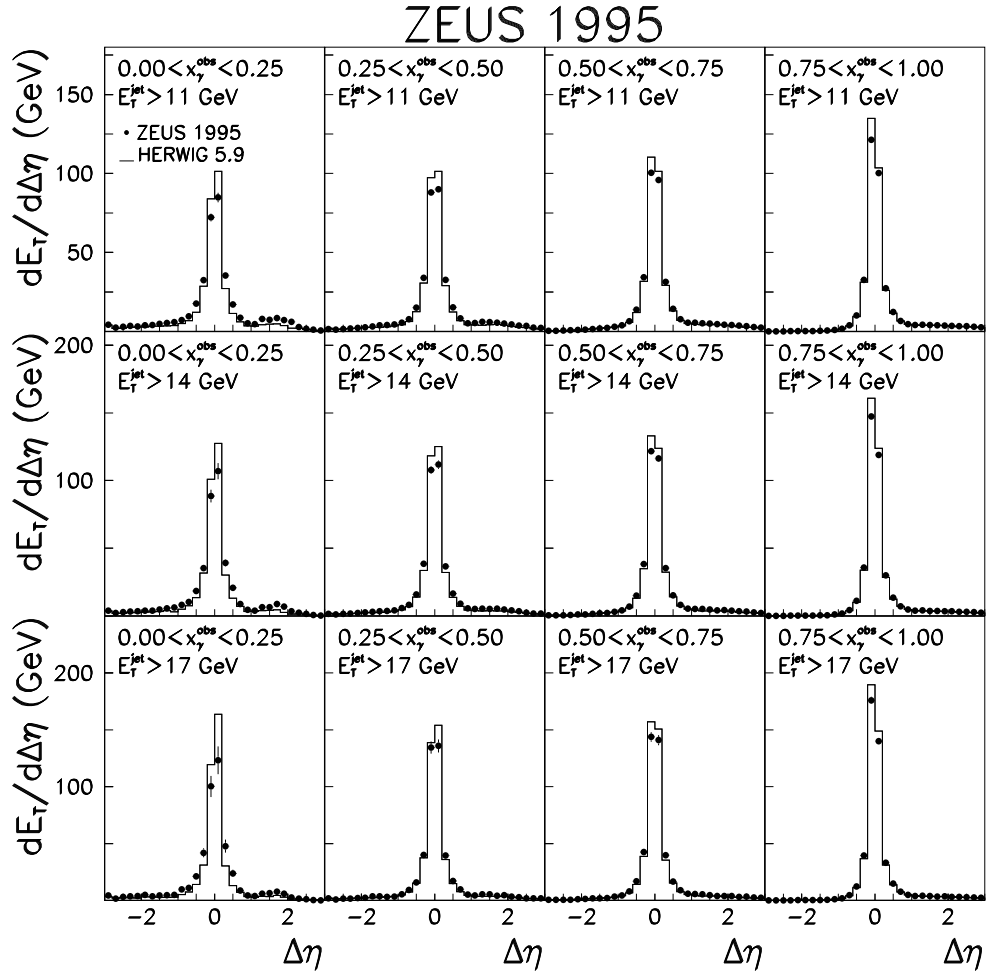}
\end{center}
\caption{\footnotesize The transverse energy flow around the jet axis (integrated over $\vert\Delta\phi\vert<1$),  in three ranges of the transverse energy of the jet and in four bins in $x_\gamma^{obs}$. The data are compared to the HERWIG 5.9 predictions. For the data only statistical uncertainties are shown.}
\label{fig:etflow}
\end{figure}

In Fig. \ref{fig:etflow} the transverse energy flow around jets  
is shown as a function of the distance in pseudorapidity $\Delta\eta$, 
with respect to the jet axis, integrated over $\Delta\phi$, between 
$\phi^{jet}-1$ and $\phi^{jet}+1$, where $\phi$ is measured in radians. 
The transverse energy 
flows are shown in bins of $E_T^{jet}$ and $x_\gamma^{obs}$. 
The jets are strongly collimated, with relatively little transverse energy 
away from the jets. Comparison to the HERWIG predictions shows generally 
good agreement. Only at low $x_\gamma^{obs}$ values is the energy flow 
outside the jet underestimated by the Monte Carlo model. Jets in the Monte 
Carlo are also found to be slightly narrower than jets in the data.

In the kinematic regime of the present analysis, Monte Carlo models 
that do not include a simulation of underlying events are able to 
describe the $x_\gamma^{obs}$ distribution and the transverse energy 
flows, the only exception being the transverse energy flows 
in the lowest $x_\gamma^{obs}$ bins, where a small discrepancy is 
observed. 
As these distributions are considered to be particularly 
sensitive to underlying events, this indicates that such processes play 
no role in the present kinematic regime. This result is different from 
what was observed in previous photoproduction 
analyses~\cite{zeus_incl1,zeus_dij1,h1_dij1,zeus_dij2,zeus_incl2}, in which 
jets at lower transverse energies were studied, where it was 
shown that the description of the data is improved when a model 
simulating soft or hard underlying events is included in the simulations.

\section{Unfolding and Systematics} \label{syst}
The unfolding of the cross section is done by multiplying 
the number of events reconstructed in each bin by a correction factor 
determined from the HERWIG Monte Carlo sample.
This correction factor is defined as the ratio of the number of events 
generated in the bin, $N_{true}$, over $N_{rec}$, the number of 
events reconstructed in the bin.
The systematic uncertainty related to the choice of Monte Carlo model 
for the unfolding is estimated by using a different Monte 
Carlo generator, PYTHIA, to determine the correction factors. 
The HERWIG and PYTHIA Monte Carlo models differ in the 
treatment of the generation of the photon spectrum, the parton showers 
and the simulation of hadronisation effects. Nevertheless, 
both Monte Carlo models give a reasonable description of the data. 

To determine systematic uncertainties in the measured cross sections, 
several variations in the event selection have been studied.
The uncertainty in the cross section due to the  
energy-scale uncertainty is estimated by raising 
and lowering all energies in the Monte Carlo simulation by 3\% simultaneously. 
In addition $y$ and the transverse jet energies are varied by 
$\pm 2\%$ separately.

The systematic uncertainty related to the energy measurement is 
correlated from bin to bin. In the cross section figures presented in 
the next section, this uncertainty 
is shown separately. All other positive 
(negative) contributions to the cross section, from systematic uncertainties, 
are added in quadrature to yield the total positive (negative) systematic uncertainty.

\section{Results} \label{results}
The dijet  photoproduction cross section is presented as a function 
of three variables: $E_{T\,leading}^{jet}$, the transverse energy of the 
leading jet, and $\eta^{jet}_1$ and $\eta^{jet}_2$, 
the pseudorapidities of the two jets.  
Statistical and systematic uncertainties, added in quadrature, 
are shown as thin error bars. Statistical uncertainties 
alone are shown as thick error bars and the uncertainty
due to the energy scale is shown as a 
shaded band.

\subsection{Cross sections for $134 < W_{\gamma p} < 277$\,GeV}
\begin{figure}
\epsfxsize = 5.in
\epsfysize = 5.in
\epsffile{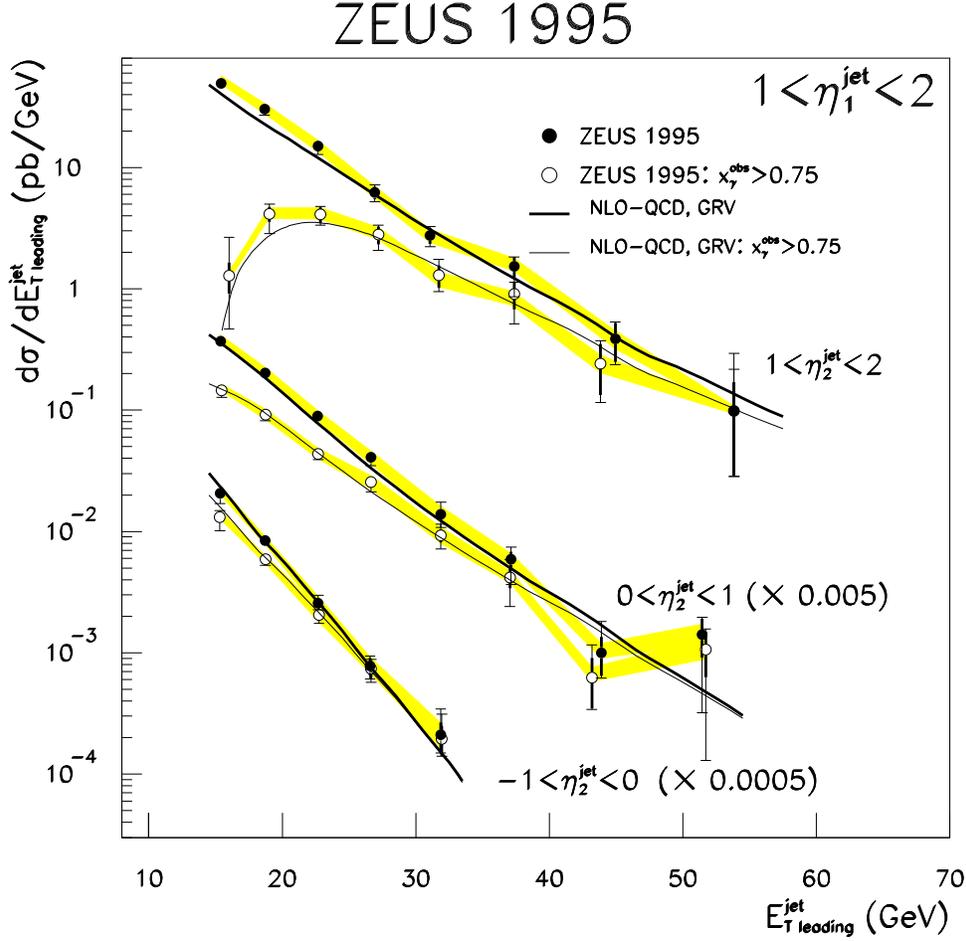}
\caption{\footnotesize Dijet cross section as a function of $E_{T\,leading}^{jet}$ for $\eta_1^{jet}$ between 1 and 2, in three regions of $\eta_2^{jet}$. The results for $-1<\eta_2^{jet}<0$ and $0<\eta_2^{jet}<1$ are scaled by the factors indicated in the figure. The filled circles correspond to the entire $x_\gamma^{obs}$ range while the open circles correspond to events with $x_\gamma^{obs}> 0.75$. The shaded band indicates the uncertainty related to the energy scale. The thick error bar indicates the statistical uncertainty and the thin error bar indicates the systematic and statistical uncertainties added in quadrature. The data are compared to NLO-QCD calculations, using the GRV-HO parameterisation for the photon structure.}
\label{fig:dsigdet1}
\end{figure}

\begin{figure}
\epsfxsize = 5.in
\epsfysize = 5.in
\epsffile{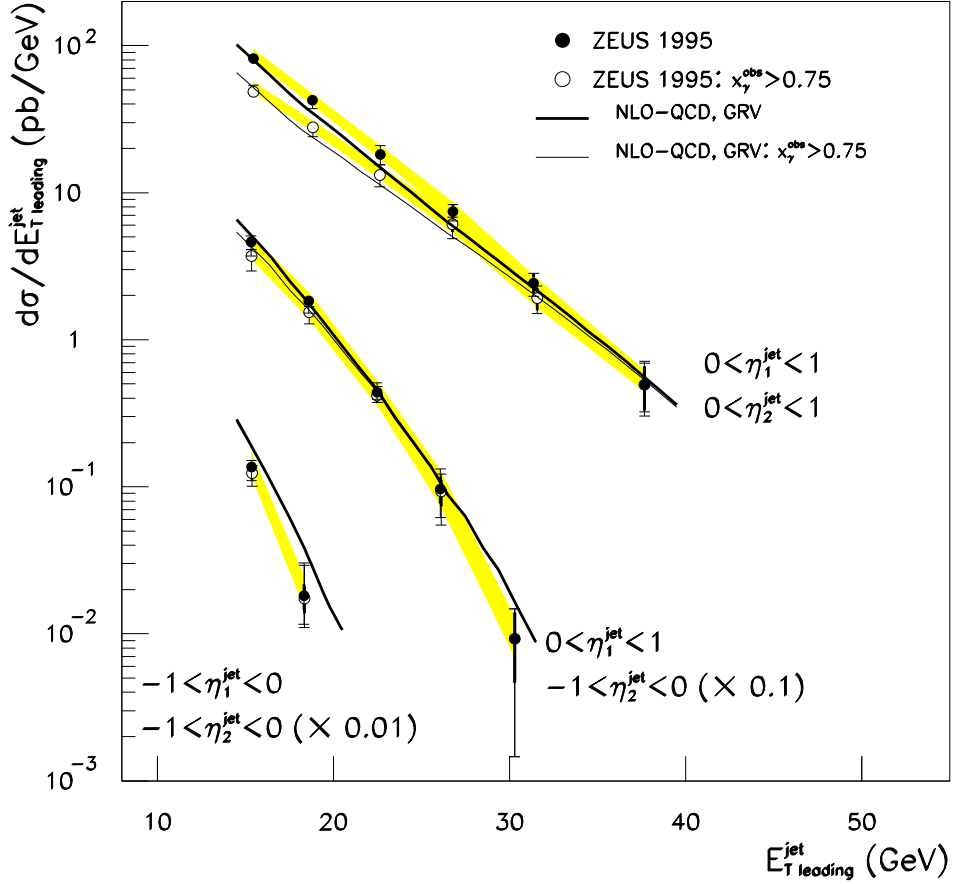}
\caption{\footnotesize Dijet cross section as a function of $E_{T\,leading}^{jet}$. For the two upper sets of data $\eta_1^{jet}$ lies between 0 and 1 and for the lower set of data  $\eta_1^{jet}$ lies between -1 and 0. The $\eta_2^{jet}$ regions are indicated  the figure. The two lower sets of data are scaled by the factors indicated in the figure. The filled circles correspond to the entire $x_\gamma^{obs}$ range while the open circles correspond to events with $x_\gamma^{obs}> 0.75$. The shaded band indicates the uncertainty related to the energy scale. The thick error bar indicates the statistical uncertainty and the thin error bar indicates the systematic and statistical uncertainties added in quadrature. The data are compared to NLO-QCD calculations, using the GRV-HO parameterisation for the photon structure. }
\label{fig:dsigdet2}
\end{figure}

The dijet cross section as a function of the transverse energy of the 
leading jet is presented for six different ranges in jet pseudorapidity. 
These cross sections have been determined 
both for the full $x_\gamma^{obs}$ range and for $x_\gamma^{obs}>0.75$. 
Numerical values for the cross sections and the uncertainties are given 
in Tab. \ref{tab:dsigdet} and 
\ref{tab:dsigdet_dir}. 
The results are plotted in Fig. \ref{fig:dsigdet1} and 
\ref{fig:dsigdet2}. 
The dijet cross section falls rapidly with increasing transverse energy 
of the leading jet. The steepest slopes occur when both jets are in 
the most backward pseudorapidity bin, $-1<\eta_{1,2}^{jet}<0$. 
High $x_\gamma^{obs}$ events dominate the cross section at backward 
angles of the jets and at high transverse energies of the jets.
This behaviour is expected on kinematic grounds, since high 
$x_\gamma^{obs}$ values give access to the highest transverse jet 
energies and to the most backward pseudorapidities.

\begin{figure}
\epsfxsize = 5.4in
\epsfysize = 5.4in
\epsffile{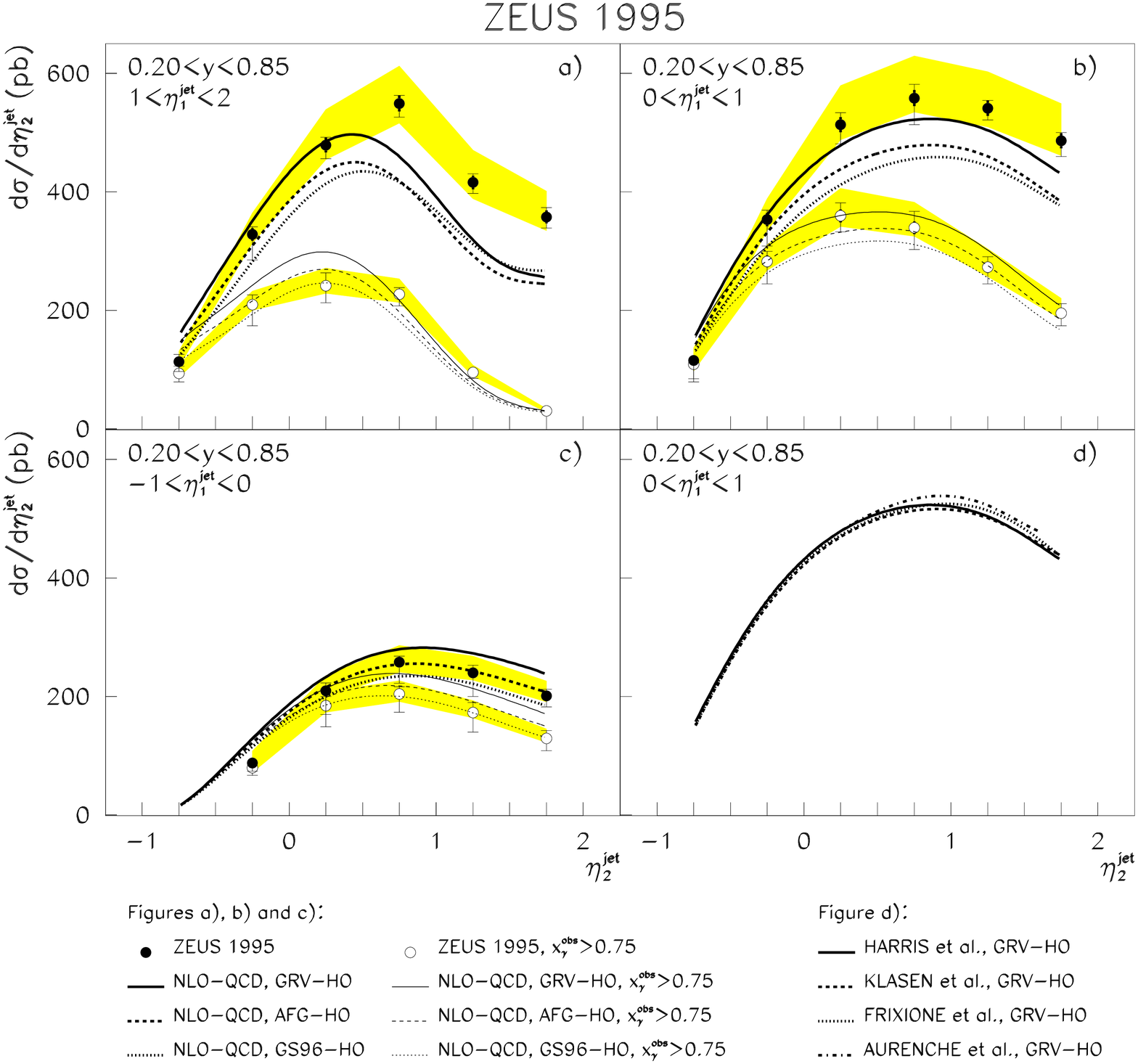}
\caption{\footnotesize Figures a), b) and c) show the dijet cross section 
as a function of $\eta_2^{jet}$ in bins of $\eta_1^{jet}$. 
The filled circles correspond to the entire $x_\gamma^{obs}$ range while the 
open circles correspond to events with $x_\gamma^{obs}> 0.75$. The shaded 
band indicates the uncertainty related to the energy scale. The thick 
error bar indicates the statistical uncertainty and the thin error bar indicates the 
systematic and statistical uncertainties added in quadrature. The full, dotted and dashed curves correspond to NLO-QCD calculations, using the GRV-HO, GS96-HO and the AFG-HO parameterisations for the photon structure, respectively. In d) the NLO-QCD results for the cross section when 
$0<\eta_1^{jet}<1$ and for a particular parameterisation of the 
photon structure are compared.}
\label{fig:dsigdeta}
\end{figure}

The data are compared to NLO-QCD calculations (see section \ref{nloqcd}).
Since the calculations from different groups are very similar, as will 
be shown in Fig. \ref{fig:dsigdeta} and \ref{fig:dsigdeta_highy}, only 
one set of calculations is shown here. This set corresponds to the 
GRV-HO~\cite{grv1,grv2} parameterisation of the photon structure, which 
gives the highest cross section. 
In general, the slopes and the absolute cross 
section are well described by the NLO-QCD calculations. 
However for events with forward jets, $1<\eta_{1,2}^{jet}<2$, 
and $E_{T\,leading}^{jet}<25$~GeV the 
data lie above the predictions (see Fig. \ref{fig:dsigdet1}) and for 
events with very backward jets, $-1<\eta_{1,2}^{jet}<0$, the measurement 
lies below the calculations (see Fig. \ref{fig:dsigdet2}). 
The Monte Carlo studies discussed in section \ref{nloqcd} show that 
fragmentation decreases the measured cross section for events with 
negative pseudorapidities. It is therefore to be expected that the 
NLO-QCD calculations, in which no parton-to-hadron fragmentation is 
included, predict a higher cross section than that observed in this region.

The dijet cross section is also presented as a function of the 
pseudorapidity of one of the jets while keeping the other jet fixed 
in specific pseudorapidity ranges. 
Numerical values for the cross section and the uncertainties are given in Tab. 
\ref{tab:dsigdeta} and \ref{tab:dsigdeta_dir} and are plotted in
Fig. \ref{fig:dsigdeta}. The cross section peaks for events with 
$\eta_2^{jet}$ near 1 and falls 
rapidly for events with $\eta_2^{jet}<0$. 
The measurements are again compared to  NLO-QCD calculations, but now 
using three different 
parameterisations for the parton densities in the photon. 
For the full $x_\gamma^{obs}$ range, at central and forward 
pseudorapidities of the jets, the data lie above all  
predictions. At backward pseudorapidities, as was the case for the 
cross section as a 
function of $E_{T\,leading}^{jet}$, the data lie 
below the calculations. In the high $x_\gamma^{obs}$ region general agreement 
is seen between the data and the predictions. 

Fig. \ref{fig:dsigdeta}d shows a comparison between the NLO-QCD 
results from four different groups for the range 
$0<\eta^{jet}_1<1$. Each calculation uses the 
same parton density distributions for the proton, CTEQ4M~\cite{cteq4m}, 
and the photon, GRV-HO~\cite{grv1,grv2}. The calculations from 
Aurenche et al., Frixione et al., Harris et al. and 
Klasen et al. agree to within a few percent.

In summary, it has been shown that NLO-QCD calculations generally 
describe the measured cross sections. However, for backward 
pseudorapidities the data are below the calculations, which is expected 
due to fragmentation, while for forward and central pseudorapidities the 
data are above the NLO predictions. In the latter kinematic region 
theoretical uncertainties are expected to be small.

\subsection{Cross sections for $212 < W_{\gamma p} < 277$~GeV}\label{highy}
The pseudorapidity dependence of the cross section has also been 
determined for events in a narrower region in $y$, which corresponds to a 
narrower range in $W_{\gamma p}$, the photon-proton CM energy.
In such a region the sensitivity to the photon structure is expected 
to be larger. This follows from the relation between $y$, $x_\gamma^{obs}$ 
and the pseudorapidities of the jets (see formula \ref{form:xg}).
Using a narrower range of $y$ values implies that the cross section 
for specific pseudorapidities of the jets corresponds to a narrower 
range of $x_\gamma^{obs}$ values. 
It is natural to select a narrow region of high $y$ values rather than a 
narrow region of low $y$ values, since, in the latter case, events with low
$x_\gamma^{obs}$ would fall out of the range of jet pseudorapidities,
$-1<\eta^{jet}<2$.

\begin{figure}
\epsfxsize = 5.4in
\epsfysize = 5.4in
\epsffile{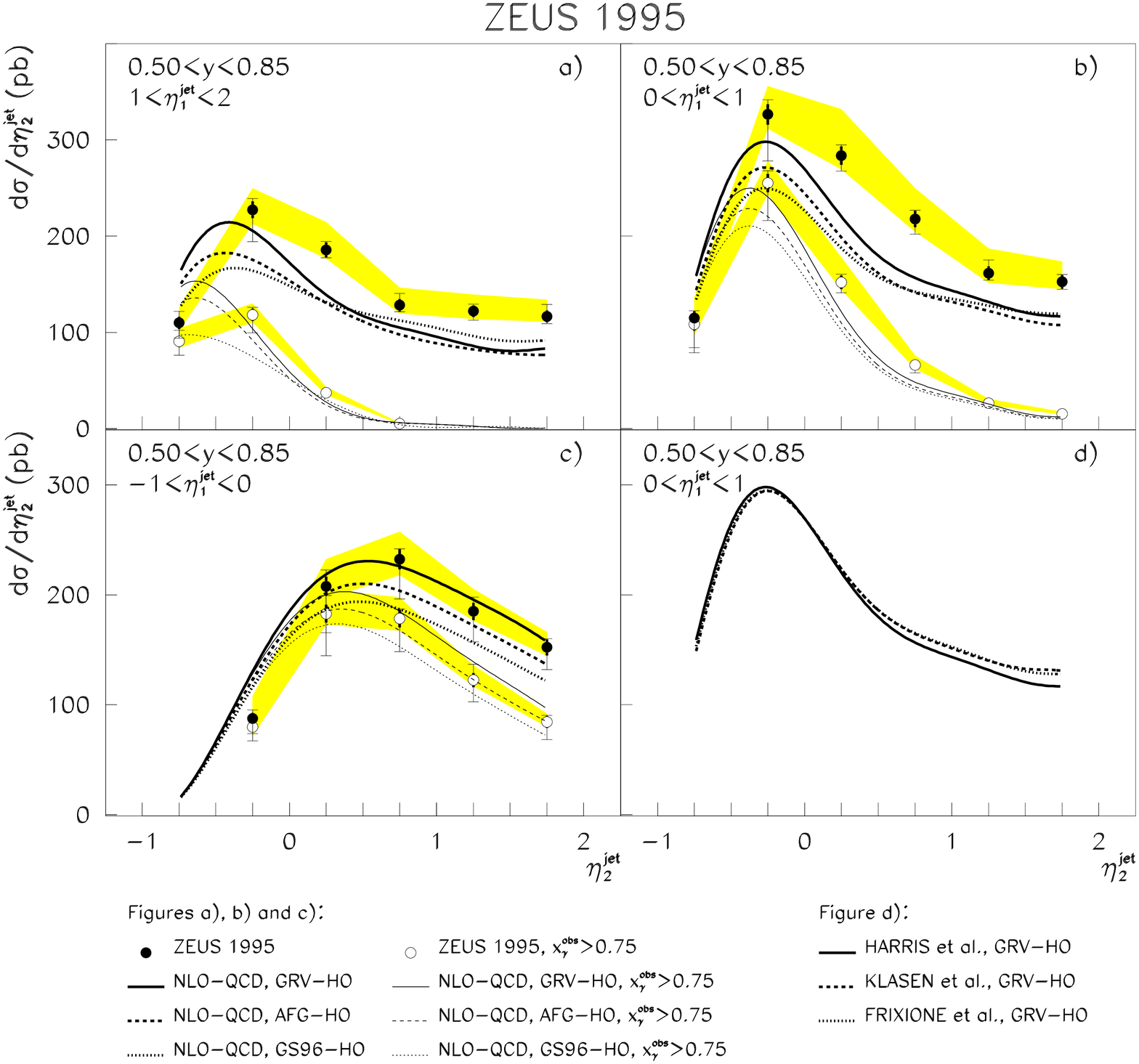}
\caption{\footnotesize Figures a), b) and c) show the dijet cross section  
as a function of $\eta_2^{jet}$ in bins of $\eta_1^{jet}$ and for $0.50<y<0.85$. The filled circles correspond to the entire $x_\gamma^{obs}$ range while the 
open circles correspond to events with $x_\gamma^{obs}> 0.75$. The shaded 
band indicates the uncertainty related to the energy scale. The thick 
error bar indicates the statistical uncertainty and the thin error bar indicates the 
systematic and statistical uncertainties added in quadrature. The full, dotted and dashed curves correspond to NLO-QCD calculations, using the GRV-HO, GS96-HO and the AFG-HO parameterisations for the photon structure, respectively. In d) the NLO-QCD results for the cross section when 
$0<\eta_1^{jet}<1$ and for a particular parameterisation of the 
photon structure are compared.}
\label{fig:dsigdeta_highy}
\end{figure}

Using a range of $0.50<y<0.85$, the cross 
section is presented as a function of the pseudorapidity of one of the 
jets while keeping the other jet fixed in a specific pseudorapidity bin.
Values for the cross section and the uncertainties are given in Tab.  
\ref{tab:dsigdeta_highy} and \ref{tab:dsigdeta_highy_dir} and are shown in 
Fig. \ref{fig:dsigdeta_highy}. The cross 
section for this high-$y$ region peaks at more backward pseudorapidities than 
the cross section for the full $y$ range, as observed in a 
previous ZEUS study~\cite{zeus_incl2}, and also the peak is more pronounced 
than for the full $y$ range. This observation is consistent with the 
expected closer correlation between $\eta^{jets}$ and $x_\gamma^{obs}$ 
when the $y$ range is restricted.
The peak in the cross sections at backward 
pseudorapidities reflects the peak near $x_\gamma^{obs}\sim 1$ 
in Fig. \ref{fig:xg} and the tail towards positive pseudorapidities 
corresponds to low $x_\gamma^{obs}$ values.

The measurements are again compared to NLO-QCD calculations using the GRV-HO, 
AFG-HO and GS96-HO parameterisations of the photon structure. 
The NLO predictions show an enhanced sensitivity to the choice of 
parameterisation for the photon structure. In particular in the region 
$1<\eta^{jet}_1<2$ there are clear differences in shape between the 
NLO predictions corresponding to different parton densities in the 
photon.
In the most backward bins, where $\eta_2^{jet}<-0.5$ or where  
$\eta_{1,2}^{jet}<0$, the data again lie below the  
calculations, but, as stated above, fragmentation effects are large 
in this region. 
At central and forward pseudorapidities, both for the full and for the 
high $x_\gamma^{obs}$ range, the data lie above the NLO calculations.

In Fig. \ref{fig:dsigdeta_highy}d a comparison is again made between 
the NLO-QCD results from different groups. The calculations agree to 
within a few percent.

The fact that the cross sections, measured in the region where jets 
are produced at central and forward pseudorapidities and where theoretical 
uncertainties are expected to be small, lie above the NLO-QCD predictions,  
suggests that in this kinematic region the 
parton densities in the photon are too small in the available 
parameterisations. 
The disagreement between the data and the calculations is observed for 
the full $x_\gamma^{obs}$ range and to a lesser extent also for 
$x_\gamma^{obs}>0.75$. It is strongest at central pseudorapidities. 
This region corresponds to values of $x_\gamma$ that lie roughly 
between 0.5 and 1.

\section{Summary and conclusions} \label{concl}

A measurement of dijet photoproduction, in the range $0.20<y<0.85$, 
$Q^2<1$~GeV$^2$, $-1<\eta^{jet}<2$, $E_{T\,leading}^{jet}>14$~GeV 
and $E_{T\,second}^{jet}>11$~GeV, has been presented.
Jets are defined in the hadronic final state by applying  
the $k_T$-clustering jet algorithm.
The cross section has been compared to NLO-QCD predictions.

For the full $y$ region, $0.20<y<0.85$, corresponding to 
$134 < W_{\gamma p} < 277$~GeV, the dijet cross section has been 
measured as a function of the transverse energy of the leading jet 
and as a function of the pseudorapidities of the jets. 
The dependence on the transverse energy of the leading jet is generally 
well described by the NLO-QCD calculations, although for events with 
two forward-going jets and $E_{T\,leading}^{jet}<25$~GeV, the data lie 
above the NLO-QCD calculations. 
Also, the cross section as a function of the 
pseudorapidities of the jets lies above the NLO-QCD calculations at 
central and forward pseudorapidities. In the region $x_\gamma^{obs}>0.75$, 
the calculations agree with the measured cross section.

In the high-$y$ region, $0.50<y<0.85$ ($212 < W_{\gamma p} < 277$~GeV), 
where a stronger sensitivity to the photon structure is expected, the cross 
section at central and forward pseudorapidities lies further above 
the predictions than for the full $y$ range. 
Also the cross section lies above 
the NLO-QCD calculations for $x_\gamma^{obs}>0.75$.

Since theoretical uncertainties are expected to be small in most of the 
kinematic regime of the present analysis, as was discussed in sections 
\ref{nloqcd} and \ref{mi}, the discrepancies observed 
between the data and the NLO-QCD calculations suggest that, in the 
kinematic region of the present analysis, the 
available parameterisations of the parton densities in the 
photon are too small. 

The results presented in this paper cover a kinematic region where both 
$x_\gamma^{obs}$ and $E_T^{jet}$, which acts as the 
factorisation scale, are high. This region has not been studied in 
$F_2^{\gamma}$ measurements.
It remains to be established whether the parton density 
functions in the photon can be modified to describe the present data 
while remaining consistent with the existing $F_2^{\gamma}$ data from 
$e^+e^-$ experiments.
It is hoped that phenomenologists carrying out comprehensive NLO-QCD fits 
will be able to include the data in this paper in their fits to determine
the parton density functions in the photon and thereby clarify this issue.

\section*{Acknowledgements}

We thank the DESY Directorate for their strong support and encouragement.
The remarkable achievements of the HERA machine group were essential
for the successful completion of this work and are greatly appreciated.
We would like to thank M. Fontannaz, S. Frixione, B. Harris and M. Klasen 
for providing us with their NLO calculations or code and for useful 
discussions on theoretical issues.

This paper was completed shortly after the tragic and untimely death of 
Prof. Dr. B.H. Wiik, Chairman of the DESY directorate. All members of the 
ZEUS collaboration wish to acknowledge the remarkable role which he 
played in the success of both the HERA project and of the ZEUS experiment. 
His inspired scientific leadership, his warm personality and his 
friendship will be sorely missed by us all.

\def\Journal#1#2#3#4{{#1} {\bf #2}, #3 (#4)}

\def\NCA{\em Nuovo Cimento}
\def\NIM{\em Nucl. Instrum. Methods}
\def\NIMA{{\em Nucl. Instrum. Methods} A}
\def\NPB{{\em Nucl. Phys.} B}
\def\PSNPB{{\em Proc. Suppl. Nucl. Phys.} B}
\def\PLB{{\em Phys. Lett.}  B}
\def\PRL{\em Phys. Rev. Lett.}
\def\PRD{{\em Phys. Rev.} D}
\def\ZPC{{\em Z. Phys.} C}
\def\EPJ{{\em Eur. Phys. J.} C}
\def\CPC{\em Comp. Phys. Commun.}

\clearpage

{\scriptsize{
\begin{table}
 \begin{center}
 \begin{tabular}{|r|r|r|r|r|}
 \hline
 \multicolumn{5}{|c|}{$d\sigma/dE_{T\,leading}^{jet}$ for: 
 $0.20<y<0.85$ and all $x_\gamma^{obs}$ values}\\
 \hline
 \multicolumn{1}{|c|}{$E_{T\,leading}^{jet}$}&
 \multicolumn{1}{|c|}{d$\sigma /$d$E_{T\,leading}^{jet}$}&
 \multicolumn{1}{c|}{$\Delta_{stat}$}&
 \multicolumn{1}{c|}{$\Delta_{syst}\;(+/-)$}&
 \multicolumn{1}{c|}{$\Delta_{E-scale}\;(+/-)$}\\
 \multicolumn{1}{|c|}{GeV}&
 \multicolumn{1}{c|}{pb/GeV}&
 \multicolumn{1}{c|}{pb/GeV}&
 \multicolumn{1}{c|}{pb/GeV}&
 \multicolumn{1}{c|}{pb/GeV}\\
 \hline
 \hline
 \multicolumn{5}{|c|}{$-1<\eta_1^{jet}<0\;\&\: -1<\eta_2^{jet}<0$}\\
 \hline
14.0 .. 17.0&  13.6&   1.2&   0.9/  -2.2&   3.2/  -1.2\\
17.0 .. 21.0&  1.81&  0.37&  1.16/ -0.54&  0.60/ -0.26\\
 \hline
 \hline
 \multicolumn{5}{|c|}{$0<\eta_1^{jet}<1\;\&\: -1<\eta_2^{jet}<0$}\\
 \hline
14.0 .. 17.0&  46.2&   1.7&   4.4/  -8.9&   4.1/  -2.1\\
17.0 .. 21.0&  18.3&   0.9&   0.7/  -3.1&   2.7/  -1.5\\
21.0 .. 25.0&   4.4&   0.4&   0.6/  -0.2&   0.8/  -0.5\\
25.0 .. 29.0&  0.97&  0.20&  0.29/ -0.29&  0.32/ -0.15\\
29.0 .. 35.0& 0.093& 0.046& 0.030/-0.063& 0.035/-0.023\\
 \hline
 \hline
 \multicolumn{5}{|c|}{$1<\eta_1^{jet}<2\;\&\: -1<\eta_2^{jet}<0$}\\
 \hline
14.0 .. 17.0&  41.2&   1.6&   2.4/  -7.1&   5.8/  -1.0\\
17.0 .. 21.0&  16.9&   0.8&   0.5/  -1.4&   1.2/  -1.2\\
21.0 .. 25.0&   5.1&   0.5&   0.7/  -0.4&   0.7/  -0.6\\
25.0 .. 29.0&  1.56&  0.26&  0.17/ -0.23&  0.29/ -0.19\\
29.0 .. 35.0&  0.42&  0.11&  0.24/ -0.05&  0.10/ -0.08\\
 \hline
 \hline
 \multicolumn{5}{|c|}{$0<\eta_1^{jet}<1\;\&\: 0<\eta_2^{jet}<1$}\\
 \hline
14.0 .. 17.0&  81.8&   3.0&   3.1/  -1.8&  11.9/  -1.6\\
17.0 .. 21.0&  42.5&   1.9&   0.5/  -4.8&   4.1/  -2.6\\
21.0 .. 25.0&  18.2&   1.2&   2.4/  -2.5&   2.6/  -1.1\\
25.0 .. 29.0&   7.5&   0.8&   0.3/  -0.7&   1.1/  -0.8\\
29.0 .. 35.0&   2.4&   0.4&   0.1/  -0.3&   0.3/  -0.3\\
35.0 .. 41.0&  0.49&  0.16&  0.14/ -0.04&  0.12/ -0.05\\
 \hline
 \hline
 \multicolumn{5}{|c|}{$1<\eta_1^{jet}<2\;\&\: 0<\eta_2^{jet}<1$}\\
 \hline
14.0 .. 17.0&  73.7&   2.0&   1.7/  -3.0&   9.0/  -0.5\\
17.0 .. 21.0&  40.4&   1.3&   0.7/  -2.6&   4.7/  -2.9\\
21.0 .. 25.0&  17.9&   0.9&   0.2/  -1.3&   1.9/  -1.6\\
25.0 .. 29.0&   8.2&   0.6&   0.1/  -1.0&   1.1/  -1.0\\
29.0 .. 35.0&   2.8&   0.3&   0.7/  -0.6&   0.4/  -0.4\\
35.0 .. 41.0&  1.18&  0.18&  0.25/ -0.41&  0.13/ -0.14\\
41.0 .. 48.0&  0.20&  0.07&  0.15/ -0.03&  0.04/ -0.02\\
48.0 .. 55.0&  0.28&  0.10&  0.04/ -0.19&  0.06/ -0.05\\
 \hline
 \hline
 \multicolumn{5}{|c|}{$1<\eta_1^{jet}<2\;\&\: 1<\eta_2^{jet}<2$}\\
 \hline
14.0 .. 17.0&  49.6&   2.3&   4.0/  -2.0&   7.1/  -0.8\\
17.0 .. 21.0&  30.4&   1.6&   1.0/  -2.9&   3.4/  -2.7\\
21.0 .. 25.0&  15.0&   1.1&   0.5/  -1.9&   1.6/  -1.5\\
25.0 .. 29.0&   6.2&   0.7&   0.7/  -0.7&   0.8/  -0.6\\
29.0 .. 35.0&   2.8&   0.4&   0.3/  -0.3&   0.3/  -0.4\\
35.0 .. 41.0&  1.53&  0.29&  0.05/ -0.60&  0.25/ -0.12\\
41.0 .. 48.0&  0.39&  0.14&  0.05/ -0.06&  0.06/ -0.07\\
48.0 .. 55.0& 0.099& 0.070& 0.183/-0.009& 0.009/-0.007\\
 \hline
 \end{tabular}
 \end{center}
\caption{{\footnotesize The dijet cross section for the full $x_\gamma^{obs}$ range and $0.20<y<0.85$, as a function of $E_{T\,leading}^{jet}$ in bins of the jet pseudorapidities.}}
\label{tab:dsigdet}
\end{table}

\begin{table}
 \begin{center}
 \begin{tabular}{|r|r|r|r|r|}
 \hline
 \multicolumn{5}{|c|}{$d\sigma/dE_{T\,leading}^{jet}$ for: 
 $0.20<y<0.85$ and $x_\gamma^{obs}>0.75$}\\
 \hline
 \multicolumn{1}{|c|}{$E_{T\,leading}^{jet}$}&
 \multicolumn{1}{|c|}{d$\sigma /$d$E_{T\,leading}^{jet}$}&
 \multicolumn{1}{c|}{$\Delta_{stat}$}&
 \multicolumn{1}{c|}{$\Delta_{syst}\;(+/-)$}&
 \multicolumn{1}{c|}{$\Delta_{E-scale}\;(+/-)$}\\
 \multicolumn{1}{|c|}{GeV}&
 \multicolumn{1}{c|}{pb/GeV}&
 \multicolumn{1}{c|}{pb/GeV}&
 \multicolumn{1}{c|}{pb/GeV}&
 \multicolumn{1}{c|}{pb/GeV}\\
 \hline
 \hline
 \multicolumn{5}{|c|}{$-1<\eta_1^{jet}<0\;\&\: -1<\eta_2^{jet}<0$}\\
 \hline
14.0 .. 17.0&  12.4&   1.2&   1.0/  -2.0&   2.9/  -1.1\\
17.0 .. 21.0&  1.74&  0.36&  1.12/ -0.52&  0.58/ -0.25\\
 \hline
 \hline
 \multicolumn{5}{|c|}{$0<\eta_1^{jet}<1\;\&\: -1<\eta_2^{jet}<0$}\\
 \hline
14.0 .. 17.0&  37.1&   1.5&   3.5/  -7.7&   3.3/  -1.7\\
17.0 .. 21.0&  15.4&   0.8&   1.1/  -2.5&   2.3/  -1.3\\
21.0 .. 25.0&   4.2&   0.4&   0.5/  -0.2&   0.8/  -0.5\\
25.0 .. 29.0&  0.93&  0.19&  0.22/ -0.32&  0.30/ -0.14\\
29.0 .. 35.0& 0.093& 0.046& 0.030/-0.063& 0.035/-0.023\\
 \hline
 \hline
 \multicolumn{5}{|c|}{$1<\eta_1^{jet}<2\;\&\: -1<\eta_2^{jet}<0$}\\
 \hline
14.0 .. 17.0&  26.2&   1.2&   3.2/  -5.9&   3.7/  -0.7\\
17.0 .. 21.0&  11.8&   0.7&   0.7/  -1.1&   0.9/  -0.9\\
21.0 .. 25.0&   4.1&   0.4&   0.6/  -0.5&   0.6/  -0.5\\
25.0 .. 29.0&  1.48&  0.25&  0.13/ -0.22&  0.27/ -0.18\\
29.0 .. 35.0&  0.39&  0.11&  0.20/ -0.02&  0.10/ -0.07\\
 \hline
 \hline
 \multicolumn{5}{|c|}{$0<\eta_1^{jet}<1\;\&\: 0<\eta_2^{jet}<1$}\\
 \hline
14.0 .. 17.0&  48.6&   2.3&   4.4/  -2.1&   7.1/  -0.9\\
17.0 .. 21.0&  27.8&   1.5&   1.6/  -3.5&   2.7/  -1.7\\
21.0 .. 25.0&  13.2&   1.0&   2.1/  -2.0&   1.9/  -0.8\\
25.0 .. 29.0&   6.1&   0.7&   0.1/  -1.0&   0.9/  -0.7\\
29.0 .. 35.0&   1.9&   0.3&   0.2/  -0.3&   0.2/  -0.3\\
35.0 .. 41.0&  0.49&  0.16&  0.11/ -0.10&  0.12/ -0.05\\
 \hline
 \hline
 \multicolumn{5}{|c|}{$1<\eta_1^{jet}<2\;\&\: 0<\eta_2^{jet}<1$}\\
 \hline
14.0 .. 17.0&  29.0&   1.3&   1.1/  -3.4&   3.6/  -0.2\\
17.0 .. 21.0&  18.4&   0.9&   1.6/  -1.9&   2.1/  -1.3\\
21.0 .. 25.0&   8.7&   0.6&   0.6/  -0.7&   0.9/  -0.8\\
25.0 .. 29.0&   5.1&   0.4&   0.1/  -0.8&   0.7/  -0.6\\
29.0 .. 35.0&  1.85&  0.23&  0.39/ -0.35&  0.30/ -0.26\\
35.0 .. 41.0&  0.83&  0.15&  0.30/ -0.32&  0.09/ -0.10\\
41.0 .. 48.0& 0.125& 0.056& 0.092/-0.012& 0.023/-0.012\\
48.0 .. 55.0&  0.21&  0.09&  0.06/ -0.16&  0.05/ -0.04\\
 \hline
 \hline
 \multicolumn{5}{|c|}{$1<\eta_1^{jet}<2\;\&\: 1<\eta_2^{jet}<2$}\\
 \hline
14.0 .. 17.0&  1.28&  0.37&  1.32/ -0.73&  0.18/ -0.02\\
17.0 .. 21.0&   4.2&   0.6&   0.6/  -1.2&   0.5/  -0.4\\
21.0 .. 25.0&   4.1&   0.6&   0.3/  -0.5&   0.4/  -0.4\\
25.0 .. 29.0&   2.8&   0.5&   0.3/  -0.5&   0.4/  -0.3\\
29.0 .. 35.0&  1.29&  0.28&  0.38/ -0.21&  0.16/ -0.16\\
35.0 .. 41.0&  0.91&  0.23&  0.05/ -0.32&  0.15/ -0.07\\
41.0 .. 48.0&  0.24&  0.11&  0.07/ -0.06&  0.04/ -0.04\\
48.0 .. 55.0& 0.099& 0.070& 0.096/-0.009& 0.009/-0.007\\
 \hline
 \end{tabular}
 \end{center}
\caption{{\footnotesize The dijet cross section for $x_\gamma^{obs}>0.75$ and $0.20<y<0.85$, as a function of $E_{T\,leading}^{jet}$ in bins of the jet pseudorapidities.}}
\label{tab:dsigdet_dir}
\end{table}

}}

{\footnotesize{
\begin{table}
 \begin{center}
 \begin{tabular}{|r|r|r|r|r|}
 \hline
 \multicolumn{5}{|c|}{$d\sigma/d\eta^{jet}$
  for: $0.20<y<0.85$ and all $x_\gamma^{obs}$ values}\\
 \hline
 \multicolumn{1}{|c|}{$\eta_2^{jet}$}&
 \multicolumn{1}{c|}{d$\sigma/$d$\eta_2^{jet}$}&
 \multicolumn{1}{c|}{$\Delta_{stat}$}&
 \multicolumn{1}{c|}{$\Delta_{syst}\;(+/-)$}&
 \multicolumn{1}{c|}{$\Delta_{E-scale}\;(+/-)$}\\
 \multicolumn{1}{|c|}{}&
 \multicolumn{1}{c|}{pb}&
 \multicolumn{1}{c|}{pb}&
 \multicolumn{1}{c|}{pb}&
 \multicolumn{1}{c|}{pb}\\
 \hline
 \hline
 \multicolumn{5}{|c|}{$-1<\eta_1^{jet}<0$}\\
 \hline
-0.5 ..  0.0&    88&     5&     6/   -13&    20/    -9\\
 0.0 ..  0.5&   209&     9&    10/   -39&    26/   -13\\
 0.5 ..  1.0&   258&     9&     4/   -34&    28/   -17\\
 1.0 ..  1.5&   240&     9&     9/   -39&    28/   -12\\
 1.5 ..  2.0&   201&     8&     7/   -17&    25/   -10\\
 \hline
 \hline
 \multicolumn{5}{|c|}{$0<\eta_1^{jet}<1$}\\
 \hline
-1.0 .. -0.5&   115&     7&     4/   -30&    21/   -12\\
-0.5 ..  0.0&   353&    11&    11/   -44&    35/   -19\\
 0.0 ..  0.5&   513&    13&    16/   -29&    66/   -25\\
 0.5 ..  1.0&   558&    14&    19/   -43&    71/   -23\\
 1.0 ..  1.5&   541&    13&     1/   -15&    61/   -32\\
 1.5 ..  2.0&   486&    13&     6/   -23&    63/   -25\\
 \hline
 \hline
 \multicolumn{5}{|c|}{$1<\eta_1^{jet}<2$}\\
 \hline
-1.0 .. -0.5&   113&     6&    10/   -15&    18/    -8\\
-0.5 ..  0.0&   328&    11&     7/   -42&    36/   -15\\
 0.0 ..  0.5&   479&    12&     4/   -19&    60/   -25\\
 0.5 ..  1.0&   549&    14&     1/   -18&    63/   -33\\
 1.0 ..  1.5&   416&    12&     9/   -14&    54/   -28\\
 1.5 ..  2.0&   358&    11&    12/   -15&    43/   -23\\
 \hline
 \end{tabular}
 \end{center}
\caption{{\footnotesize The dijet cross section, for all $x_\gamma^{obs}$ values and $0.20<y<0.85$, 
as a function of $\eta_2^{jet}$, for $\eta_1^{jet}$ fixed.}}
\label{tab:dsigdeta}
\end{table}

\begin{table}
 \begin{center}
 \begin{tabular}{|r|r|r|r|r|}
 \hline
 \multicolumn{5}{|c|}{$d\sigma/d\eta^{jet}$
  for: $0.20<y<0.85$ and $x_\gamma^{obs}>0.75$}\\
 \hline
 \multicolumn{1}{|c|}{$\eta_2^{jet}$}&
 \multicolumn{1}{c|}{d$\sigma/$d$\eta_2^{jet}$}&
 \multicolumn{1}{c|}{$\Delta_{stat}$}&
 \multicolumn{1}{c|}{$\Delta_{syst}\;(+/-)$}&
 \multicolumn{1}{c|}{$\Delta_{E-scale}\;(+/-)$}\\
 \multicolumn{1}{|c|}{}&
 \multicolumn{1}{c|}{pb}&
 \multicolumn{1}{c|}{pb}&
 \multicolumn{1}{c|}{pb}&
 \multicolumn{1}{c|}{pb}\\
 \hline
 \hline
 \multicolumn{5}{|c|}{$-1<\eta_1^{jet}<0$}\\
 \hline
-0.5 ..  0.0&    80&     5&     8/   -12&    18/    -8\\
 0.0 ..  0.5&   185&     8&    13/   -35&    23/   -11\\
 0.5 ..  1.0&   204&     8&     9/   -30&    23/   -13\\
 1.0 ..  1.5&   173&     8&    15/   -32&    20/    -9\\
 1.5 ..  2.0&   129&     7&    12/   -19&    16/    -7\\
 \hline
 \hline
 \multicolumn{5}{|c|}{$0<\eta_1^{jet}<1$}\\
 \hline
-1.0 .. -0.5&   109&     7&     3/   -29&    20/   -11\\
-0.5 ..  0.0&   283&    10&    14/   -37&    28/   -15\\
 0.0 ..  0.5&   359&    11&    19/   -25&    46/   -18\\
 0.5 ..  1.0&   339&    11&    26/   -35&    43/   -14\\
 1.0 ..  1.5&   273&     9&    15/   -27&    31/   -16\\
 1.5 ..  2.0&   195&     8&    14/   -20&    25/   -10\\
 \hline
 \hline
 \multicolumn{5}{|c|}{$1<\eta_1^{jet}<2$}\\
 \hline
-1.0 .. -0.5&    94&     6&    11/   -13&    15/    -7\\
-0.5 ..  0.0&   210&     8&    14/   -35&    23/    -9\\
 0.0 ..  0.5&   241&     9&    20/   -27&    30/   -12\\
 0.5 ..  1.0&   227&     9&     7/   -18&    26/   -14\\
 1.0 ..  1.5&    95&     6&     6/    -8&    12/    -6\\
 1.5 ..  2.0&    30&     3&     2/    -2&     4/    -2\\
 \hline
 \end{tabular}
 \end{center}
\caption{{\footnotesize The dijet cross section for $x_\gamma^{obs}>0.75$ and $0.20<y<0.85$, as a function of $\eta_2^{jet}$, for $\eta_1^{jet}$ fixed.}}
\label{tab:dsigdeta_dir}
\end{table}
}}

{\footnotesize{
\begin{table}
 \begin{center}
 \begin{tabular}{|r|r|r|r|r|}
 \hline
 \multicolumn{5}{|c|}{$d\sigma/d\eta^{jet}$
  for: $0.50<y<0.85$ and all $x_\gamma^{obs}$ values}\\
 \hline
 \multicolumn{1}{|c|}{$\eta_2^{jet}$}&
 \multicolumn{1}{c|}{d$\sigma/$d$\eta_2^{jet}$}&
 \multicolumn{1}{c|}{$\Delta_{stat}$}&
 \multicolumn{1}{c|}{$\Delta_{syst}\;(+/-)$}&
 \multicolumn{1}{c|}{$\Delta_{E-scale}\;(+/-)$}\\
 \multicolumn{1}{|c|}{}&
 \multicolumn{1}{c|}{pb}&
 \multicolumn{1}{c|}{pb}&
 \multicolumn{1}{c|}{pb}&
 \multicolumn{1}{c|}{pb}\\
 \hline
 \hline
 \multicolumn{5}{|c|}{$-1<\eta_1^{jet}<0$}\\
 \hline
-0.5 ..  0.0&    88&     5&     6/   -13&    20/    -9\\
 0.0 ..  0.5&   208&     9&    12/   -41&    25/   -11\\
 0.5 ..  1.0&   232&     9&     2/   -35&    25/   -14\\
 1.0 ..  1.5&   185&     8&    11/   -26&    20/    -8\\
 1.5 ..  2.0&   152&     7&     2/   -19&    14/    -8\\
 \hline
 \hline
 \multicolumn{5}{|c|}{$0<\eta_1^{jet}<1$}\\
 \hline
-1.0 .. -0.5&   115&     7&     4/   -30&    21/   -12\\
-0.5 ..  0.0&   326&    11&    11/   -47&    29/   -15\\
 0.0 ..  0.5&   284&    10&     6/   -13&    48/   -14\\
 0.5 ..  1.0&   218&     9&     3/   -13&    32/   -13\\
 1.0 ..  1.5&   162&     7&    12/    -2&    26/   -10\\
 1.5 ..  2.0&   153&     7&     3/    -3&    21/    -8\\
 \hline
 \hline
 \multicolumn{5}{|c|}{$1<\eta_1^{jet}<2$}\\
 \hline
-1.0 .. -0.5&   110&     6&    10/   -15&    16/    -8\\
-0.5 ..  0.0&   227&     9&     8/   -32&    22/   -14\\
 0.0 ..  0.5&   186&     8&     4/    -4&    29/    -9\\
 0.5 ..  1.0&   128&     6&    11/    -3&    18/    -8\\
 1.0 ..  1.5&   122&     6&     4/    -7&    17/    -8\\
 1.5 ..  2.0&   117&     6&    11/    -5&    17/    -5\\
 \hline
 \end{tabular}
 \end{center}
\caption{{\footnotesize The dijet cross section, for all $x_\gamma^{obs}$ values and $0.50<y<0.85$, 
as a function of $\eta_2^{jet}$, for $\eta_1^{jet}$ fixed.}}
\label{tab:dsigdeta_highy}
\end{table}

\begin{table}
 \begin{center}
 \begin{tabular}{|r|r|r|r|r|}
 \hline
 \multicolumn{5}{|c|}{$d\sigma/d\eta^{jet}$
  for: $0.50<y<0.85$ and $x_\gamma^{obs}>0.75$}\\
 \hline
 \multicolumn{1}{|c|}{$\eta_2^{jet}$}&
 \multicolumn{1}{c|}{d$\sigma/$d$\eta_2^{jet}$}&
 \multicolumn{1}{c|}{$\Delta_{stat}$}&
 \multicolumn{1}{c|}{$\Delta_{syst}\;(+/-)$}&
 \multicolumn{1}{c|}{$\Delta_{E-scale}\;(+/-)$}\\
 \multicolumn{1}{|c|}{}&
 \multicolumn{1}{c|}{pb}&
 \multicolumn{1}{c|}{pb}&
 \multicolumn{1}{c|}{pb}&
 \multicolumn{1}{c|}{pb}\\
 \hline
 \hline
 \multicolumn{5}{|c|}{$-1<\eta_1^{jet}<0$}\\
 \hline
-0.5 ..  0.0&    80&     5&     8/   -12&    18/    -8\\
 0.0 ..  0.5&   183&     8&    15/   -37&    22/   -10\\
 0.5 ..  1.0&   178&     8&     4/   -29&    19/   -11\\
 1.0 ..  1.5&   122&     6&    13/   -18&    13/    -5\\
 1.5 ..  2.0&    84&     5&     3/   -15&     8/    -4\\
 \hline
 \hline
 \multicolumn{5}{|c|}{$0<\eta_1^{jet}<1$}\\
 \hline
-1.0 .. -0.5&   108&     7&     4/   -29&    20/   -11\\
-0.5 ..  0.0&   255&     9&     9/   -38&    23/   -12\\
 0.0 ..  0.5&   152&     7&     5/    -9&    26/    -8\\
 0.5 ..  1.0&    66&     5&     0/    -7&    10/    -4\\
 1.0 ..  1.5&    27&     3&     2/    -1&     4/    -2\\
 1.5 ..  2.0&    15&     2&     2/    -1&     2/    -1\\
 \hline
 \hline
 \multicolumn{5}{|c|}{$1<\eta_1^{jet}<2$}\\
 \hline
-1.0 .. -0.5&    91&     6&    10/   -13&    13/    -6\\
-0.5 ..  0.0&   118&     6&     5/   -18&    12/    -7\\
 0.0 ..  0.5&    37&     3&     0/    -4&     6/    -2\\
 0.5 ..  1.0&   5.0&   1.2&   2.5/  -0.5&   0.7/  -0.3\\
 \hline
 \end{tabular}
 \end{center}
\caption{{\footnotesize The dijet cross section for $x_\gamma^{obs}>0.75$ and $0.50<y<0.85$, as a function of $\eta_2^{jet}$, for $\eta_1^{jet}$ fixed.}}
\label{tab:dsigdeta_highy_dir}
\end{table}
}}

\end{document}